\documentclass[10pt,twocolumn,secnumarabic, nobibnotes, aps, prd,superscriptaddress,showkeys]{revtex4-2}
\usepackage{xcolor,soul}
\usepackage{braket}
\usepackage{amsmath}
\usepackage{wrapfig}
\usepackage{multirow} 
\usepackage{physics}
\usepackage{lipsum}
\usepackage{subfigure}
\usepackage{wrapfig}
\usepackage{xparse,xcoffins}
\usepackage{graphicx}
\usepackage{comment}
\sethlcolor{red}

\ExplSyntaxOn
\NewCoffin\imagecoffin
\NewCoffin\labelcoffin
\keys_define:nn { miguel/label }
 {
  label   .tl_set:N = \l_miguel_label_tl,
  labelbox .bool_set:N = \l_miguel_label_box_bool,
  labelbox .default:n = true,
  fontsize .tl_set:N = \l_miguel_label_size_tl,
  fontsize .initial:n = \footnotesize,
  pos .choice:,
  pos/nw .code:n = \tl_set:Nn \l_miguel_label_pos_tl { left,up },
  pos/ne .code:n = \tl_set:Nn \l_miguel_label_pos_tl { right,up },
  pos/sw .code:n = \tl_set:Nn \l_miguel_label_pos_tl { left,down },
  pos/se .code:n = \tl_set:Nn \l_miguel_label_pos_tl { right,down },
  pos/n .code:n = \tl_set:Nn \l_miguel_label_pos_tl { hc,up },
  pos/w .code:n = \tl_set:Nn \l_miguel_label_pos_tl { left,vc },
  pos/s .code:n = \tl_set:Nn \l_miguel_label_pos_tl { hc,down },
  pos/e .code:n = \tl_set:Nn \l_miguel_label_pos_tl { right,vc },
  pos .initial:n = nw,
  unknown .code:n   = \clist_put_right:Nx \l_miguel_label_clist
                       { \l_keys_key_tl = \exp_not:n { #1 } }
 }
\clist_new:N \l_miguel_label_clist
\box_new:N \l_miguel_label_box
\box_new:N \l_miguel_label_image_box
\NewDocumentCommand{\xincludegraphics}{O{}m}
 {
  \group_begin:
  \tl_clear:N \l_miguel_label_tl
  \clist_clear:N \l_miguel_label_clist
  \keys_set:nn { miguel/label } { #1 }
  \tl_if_empty:NTF \l_miguel_label_tl
   {
    \miguel_includegraphics:Vn \l_miguel_label_clist { #2 }
   }
   {
    \SetHorizontalCoffin\imagecoffin
     {
      \miguel_includegraphics:Vn \l_miguel_label_clist { #2 }
     }
    \SetHorizontalCoffin\labelcoffin
     {
      \raisebox{\depth}
       {
        \bool_if:NTF \l_miguel_label_box_bool
         { \fcolorbox{white}{white}{\l_miguel_label_size_tl\l_miguel_label_tl} }
         { \l_miguel_label_size_tl\l_miguel_label_tl }
       }
     }
    \SetVerticalPole\imagecoffin{left}{3pt+\CoffinWidth\labelcoffin/2}
    \SetVerticalPole\imagecoffin{right}{\Width-3pt-\CoffinWidth\labelcoffin/2}
    \SetHorizontalPole\imagecoffin{up}{\Height-3pt-\CoffinHeight\labelcoffin/2}
    \SetHorizontalPole\imagecoffin{down}{3pt+\CoffinHeight\labelcoffin/2}
    \use:x{\JoinCoffins\imagecoffin[\l_miguel_label_pos_tl]\labelcoffin[vc,hc]}
    \TypesetCoffin\imagecoffin
   }
   \group_end:
 }
\NewDocumentCommand{\setlabel}{m}
 {
  \keys_set:nn { miguel/label } { #1 }
 }
\cs_new_protected:Nn \miguel_includegraphics:nn
 {
  \includegraphics[#1]{#2}
 }
\cs_generate_variant:Nn \miguel_includegraphics:nn { V }
\ExplSyntaxOff
\makeatletter
\newcommand{\twocolumncaption}{\@dblarg\@twocolumncaption}
\def\@twocolumncaption[#1]#2{  \renewcommand{\@makecaption}[2]{    \par\vskip\abovecaptionskip\begingroup\small\rmfamily
    \splittopskip=0pt
    \setbox\@tempboxa=\vbox{
      \@arrayparboxrestore \let \\\@normalcr
      \hsize=.5\hsize \advance\hsize-1em
      \let\\\heading@cr
      \noindent ##1\ ##2\par    }    \vbadness=10000
    \setbox\z@=\vsplit\@tempboxa to .55\ht\@tempboxa
    \setbox\z@=\vtop{\hrule height 0pt \unvbox\z@}
    \setbox\tw@=\vtop{\hrule height 0pt \unvbox\@tempboxa}
    \noindent\box\z@\hfill\box\tw@\par
    \endgroup\vskip \belowcaptionskip
  }  \setlength{\abovecaptionskip}{4ex}  \caption[#1]{#2}}
\begin{document}

\title{Quantum Circuit Benchmarking on IBM Brisbane: Performance Insights from Superconducting Qubit Models}

\author{J. Thirunirai Selvam}
\author{S. Saravana Veni}
    \email{s_saravanaveni@cb.amrita.edu }
    \affiliation{Department of Physics, Amrita School of Physical Sciences,
 Amrita Vishwa Vidyapeetham, Coimbatore, 641112, Tamil Nadu, India
}
\begin{abstract}

This paper investigates quantum communication using superconducting qubits, emphasizing the simulation and control of quantum systems via IBM’s Brisbane quantum processor. We focus on implementing fundamental quantum gates and analyzing the evolution of entangled states, which are essential for secure and reliable information transfer. The study highlights the role of entanglement as a critical resource in quantum communication, enabling secure connectivity across quantum networks. Simulations incorporate realistic conditions, including decoherence and noise, to assess the practical viability of entangled-state operations. Additionally, we explore the extension of these systems to simulate key quantum models such as the Jaynes–Cummings and longitudinal Ising models, offering insight into complex interactions in superconducting architectures. The findings advance quantum information science by demonstrating the potential of superconducting qubit systems for both foundational research and real-world applications in quantum communication.
\end{abstract}


%\titlepgskip=-15pt

\maketitle
\section{Introduction}
\vspace{1em}

Quantum computing employs the principles of quantum physics to perform calculations unattainable by traditional systems \cite{b1}. Quantum processing fundamentally relies on quantum bits, or qubits, which—unlike conventional bits—can exist in superpositions and become entangled. These distinctive quantum characteristics offer powerful computational frameworks, enabling exponential acceleration in tasks such as factoring, optimization and modeling quantum systems. Among the various physical implementations of qubits, superconducting qubits have emerged as a leading platform due to their scalability, compatibility with semiconductor fabrication processes and rapid gate speeds \cite{b2,b3,b4}. Superconducting qubits are constructed using nonlinear, non-dissipative circuits with Josephson junctions, elements that enable the system to exhibit quantized energy levels. When cooled to cryogenic temperatures, typically below 20 millikelvin, these circuits function as synthetic atoms \cite{b5}. The most widely used superconducting qubit is the transmon, a variant of the charge qubit designed to reduce sensitivity to charge noise \cite{b6,b7}. It operates within an anharmonic potential well resulting from the interplay between Josephson energy and charging energy \cite{b8}. The first two energy levels of this system form the computational basis states, denoted $\ket{0}$ and $\ket{1}$ \cite{b9}. Entangling gates between qubits are implemented via calibrated couplings and logical operations are executed using microwave pulses \cite{b10}. Major technology firms such as Google, IBM and Rigetti have developed quantum processors containing hundreds of superconducting qubits. These devices support small-scale quantum computations and experiments \cite{b11,b12}. Presently, these systems operate within the Noisy Intermediate-Scale Quantum (NISQ) era, marked by experimental validation and algorithmic development, despite the lack of full error correction capabilities \cite{b13,b14}. Quantum systems are susceptible to a wide range of noise sources, both intrinsic and external. For example, cosmic rays may delay transitions between energy levels \cite{b15}. Quantum error correction is therefore a critical area of research, with advanced error-correcting codes offering potential solutions \cite{b16}. Stabilizing quantum states remains experimentally challenging and the mitigation of noise and delays depends heavily on the choice of superconducting materials \cite{b17}. Recent advances in quantum computing include the December 2023 release of IBM’s Condor processor, which supports up to 156 qubits with low error rates. This device aims to enhance fidelity over earlier models \cite{b18, b19}. IBM has integrated these processors into platforms such as Eagle and Condor, available through open-source and subscription services for academic and industrial use \cite{b20, b21}. However, increasing qubit counts tend to raise error probabilities due to reduced fidelity.

Quantum computing has numerous practical applications. For instance, trapped ions use electromagnetic forces to generate entanglement with fidelities around 90\% \cite{b22}, while quantum dots employ semiconductor confinement effects for qubit formation.An IBM quantum simulation utilized Suzuki–Trotter decomposition to study nearest-neighbor interactions among superconducting qubits for the Heisenberg, XY model and Ising models. The fidelity of simulations varied with iteration  and experimental density matrices deviated from theoretical predictions over time. This study showcased quantum state evolution and underscored the potential of quantum computers to simulate complex systems \cite{b23}.

This work investigates the principles design  and challenges of superconducting qubit-based quantum computing. It focuses on the underlying physics, gate implementations, qubit architectures and efforts to address limitations related to scalability and decoherence. The primary goal is to deepen the understanding of quantum communication by simulating and analyzing superconducting qubit systems. Specifically, the study explores foundational theoretical models,such as the Jaynes–Cummings model and the longitudinal Ising model,in the context of secure communication. This is achieved through gate-level simulations and entanglement protocols implemented on IBM’s quantum hardware. Superconducting qubits have achieved gate fidelities of up to 95\% and institutions including IBM, Google, Willow and the Sycamore project continue working to improve performance and reduce quantum error correction overhead. These advancements are steadily increasing the reliability and accuracy of quantum computing systems.

The integration of superconductors and quantum computing yields "superconducting qubits". Josephson junctions within qubits help preserve superposition and coherence, while entanglement extends quantum behavior. Superconducting qubits are commonly classified into:
 Transmon Qubits \cite{b24} , 
Flux Qubits \cite{b25}  and 
Phase Qubits \cite{b26}. This study primarily focuses on the transmon qubit due to its superior coherence times. Leveraging transmon architectures enhances quantum performance, thereby improving information extraction in practical applications.

\section{Hamiltonian of Cooper-Pair }
\vspace{2em}
To comprehend the dynamics of a Cooper pair \cite{b27}, we examine a particle of mass $m$ traversing one dimension subjected to a potential V(x).  Within the realm of quantum physics, the total energy of this particle is denoted by the Hamiltonian operator \( \hat{H} \), which includes both kinetic and potential energy components.

\begin{equation}
\hat{H} = -\frac{\hbar^2}{2m} \frac{d^2}{dx^2} + V(x)
\end{equation}

Equation~(1) represents the time-independent Schrödinger equation~\cite{b28}, a fundamental equation in quantum mechanics that governs the energy and spatial behavior of a quantum particle in a potential $V(x)$. The Hamiltonian operator $\hat{H}$ comprises two key components are kinetic energy term and the potential energy term. The kinetic energy term, which involves Planck’s reduced constant \( \hbar \) and the particle’s mass $m$, encapsulates the wave-like nature of the particle through the second spatial derivative of the wavefunction. This term describes how the particle propagates and exhibits interference patterns. The potential energy term $ V(x)$, in contrast, accounts for the external spatially dependent influences acting on the particle, such as electromagnetic fields or confining potentials. When the potential V(x) is periodic, i.e., $V(x + a) = V(x)$ for some lattice constant $a$, Bloch’s theorem becomes applicable~\cite{b29}. According to this theorem, the solutions of the Schrödinger equation can be expressed as plane waves modulated by functions that share the same periodicity as the potential. This insight is particularly important in the study of electrons in crystalline solids, where periodic potentials lead to the formation of energy bands.

\begin{equation}
  \psi_k(x) = e^{ikx} u_k(x),  
\end{equation}

where $u_k(x)$ has the same periodicity as the potential. This form, shown in Equation (2), reflects the translational symmetry of the system and is essential for analyzing electrons in crystals and periodic structures 
 $ T $.

Applying the kinetic energy operator to $ \psi_k(x) $ using the product rule introduces terms involving the wave vector $ k $. This results the modified Schrödinger equation

\begin{equation}
    \left[ -\frac{\hbar^2}{2m} \left( \frac{d}{dx} + ik \right)^2 + V(x) \right] u_k(x) = E_k u_k(x).
\end{equation}
Eqn. (3) gives the Hamiltonian acts on the periodic part of the wavefunction, \( u_k(x) \), with the momentum operator effectively shifted by \( \hbar k \). This shift reflects the influence of the crystal momentum \( k \) on the energy spectrum, a key characteristic of electrons in periodic potentials, as described by Bloch’s theorem.
\begin{equation}
H = \frac{\hbar^2}{2m} \left( -i \frac{d}{dx} + k \right)^2 + V(x)
\label{eq:bloch_hamiltonian}
\end{equation}
Eqn. (4), results the Hamiltonian operator on \( u_k(x) \)  and the momentum operator is shifted by \( \hbar k \), show the effect of the wave vector \( k \) on the kinetic energy. In superconducting circuits, energy is stored through electrostatic imbalance rather than momentum. When there is a difference between the number of Cooper pairs (i.e., quantized charge) and the charge induced by an externally applied gate voltage, the system stores energy in the form of charging energy.
\begin{equation}
H_C = 4E_C (n - n_g)^2
\label{eq:charging_energy}
\end{equation}
Here, \( E_C \) is the charging energy scale, determined by the capacitance via \( E_C = \frac{e^2}{2C} \). The quantity \( n \) is the number of excess Cooper pairs , while \( n_g \) is the gate-induced charge, controlled continuously by an external voltage~\cite{b30}. Eqn.(5)  emphasizes that the energy increases quadratically with the deviation between the actual number of Cooper pairs and the gate-induced preferred charge. Another critical energy contribution in superconducting circuits arises from Josephson tunneling, where Cooper pairs coherently tunnel across a Josephson junction, which described by the Josephson energy,

\begin{equation}
H_J = -E_J \cos(\phi)
\label{eq:josephson_energy}
\end{equation}

Here,  \( E_J \)  represents the Josephson coupling energy, whereas \( \phi \) denotes the superconducting phase difference across the junction.  Equation (6) delineates the Josephson energy, which characterizes the tunneling of Cooper pairs across the junction and is clearly contingent upon the phase difference between the superconductors \cite{b31}.  The total Hamiltonian of a charge qubit, shown by the Cooper-pair box, is derived from the sum of the charging energy and the Josephson energy.

\begin{equation}
\hat{H}_{cp} = 4E_C(n - n_g)^2 - E_J \cos(\phi)
\end{equation}

The second term is the Josephson energy, which depends on the superconducting phase difference \( \phi \) across the junction.

\section{Hamiltonian of Jaynes-Cumming Model }
\vspace{2em}
The quantum dynamics of the Cooper-pair box are governed by the interplay between these two energies, a balance that is crucial for the qubit's ability to store and manipulate quantum information.

\begin{equation}
    \Gamma = \frac{\kappa g^2}{\Delta^2},
\end{equation}
where \( g \) is the coupling strength, \( \Delta \) is the detuning and \( \kappa \) is the photon decay rate of the cavity. Equation (8) typically represents the interaction Hamiltonian or the energy shift that occurs when the qubit is coupled to a resonant cavity. 

The photon decay rate of a cavity is given by:

\begin{equation}
\kappa = \frac{\omega_r}{Q_r}
\end{equation}

Equation (9), results in an effective decay rate for the qubit , where \( \omega_r \) is the resonant frequency of the cavity and \( Q_r \) is its quality factor. This relationship illustrates how a qubit, even when well-isolated, can experience decay through its coupling to a lossy resonator.The design and execution of quantum information processing systems heavily relies on these decay mechanisms, which are basic concepts in circuit quantum electrodynamics (circuit QED).

The Purcell effect explains how a quantum emitter's (such a qubit's) spontaneous emission rate changes when it is connected to a resonant cavity \cite{b32}. To get the Purcell decay rate, we begin with the Jaynes–Cummings Hamiltonian.  A two-level system (qubit) interacting with a single-mode cavity is represented by this Hamiltonian.

The interaction between a single mode of a quantized electromagnetic field and a two-level atom (qubit) is articulated by the Jaynes–Cummings model \cite{b33}. This model elucidates the energy exchange between the atom and the field through the rotating wave approximation, resulting in phenomena such as Rabi oscillations and vacuum Rabi splitting \cite{b34}. It is fundamental in quantum optics and extensively employed in cavity and circuit quantum electrodynamics, particularly in quantum computing with superconducting qubits \cite{b35}.

\begin{equation}
    H_{JC} = \frac{\hbar \omega_q}{2} \sigma_z + \hbar \omega_r a^\dagger a + \hbar g (\sigma^+ a + \sigma^- a^\dagger)
\end{equation}

It contains three primary elements. The first term gives how much energy is kept in the cavity, where quantum operators count the quantity of photons and their frequency \cite{b36}. The second term talks about the qubit's energy, which can be in one of two states either ground or excited. There is a certain frequency that separates these two states. In the third portion, we can see how the qubit and the cavity trade energy. The qubit either sends a photon into the cavity or takes one from it. Transmon qubits aim to stabilize the qubit and improve operational speed.  Isaac Rabi empirically identified this phenomena and received the Nobel Prize in Physics in 1944 for his discoveries.

The Rabi frequency is crucial for manipulating qubit populations in superconducting qubits.   It clarifies the frequency transition between phases in quantum systems, enabling the phase gate operation. The quantum circuit shown in fig. 1 is a series of gate operations on multiple qubits, starting with single-qubit rotations using \( R_z \) and \( \sqrt{X} \) gates, followed by Hadamard gates and entangling operations, likely CNOTs and concluding with measurement in the Z basis. These operations reflect experimental realizations of theoretical models like the Jaynes–Cummings (JC) interaction, where microwave pulses precisely control the quantum state evolution. The circuit is likely designed for process tomography or fidelity estimation, making use of superposition, entanglement and measurement to evaluate the performance of quantum gates. Overall, the JC model provides the physical mechanism for implementing and analyzing quantum logic operations in superconducting qubit systems.

\begin{figure*}
\centering
\includegraphics[width=1\linewidth]{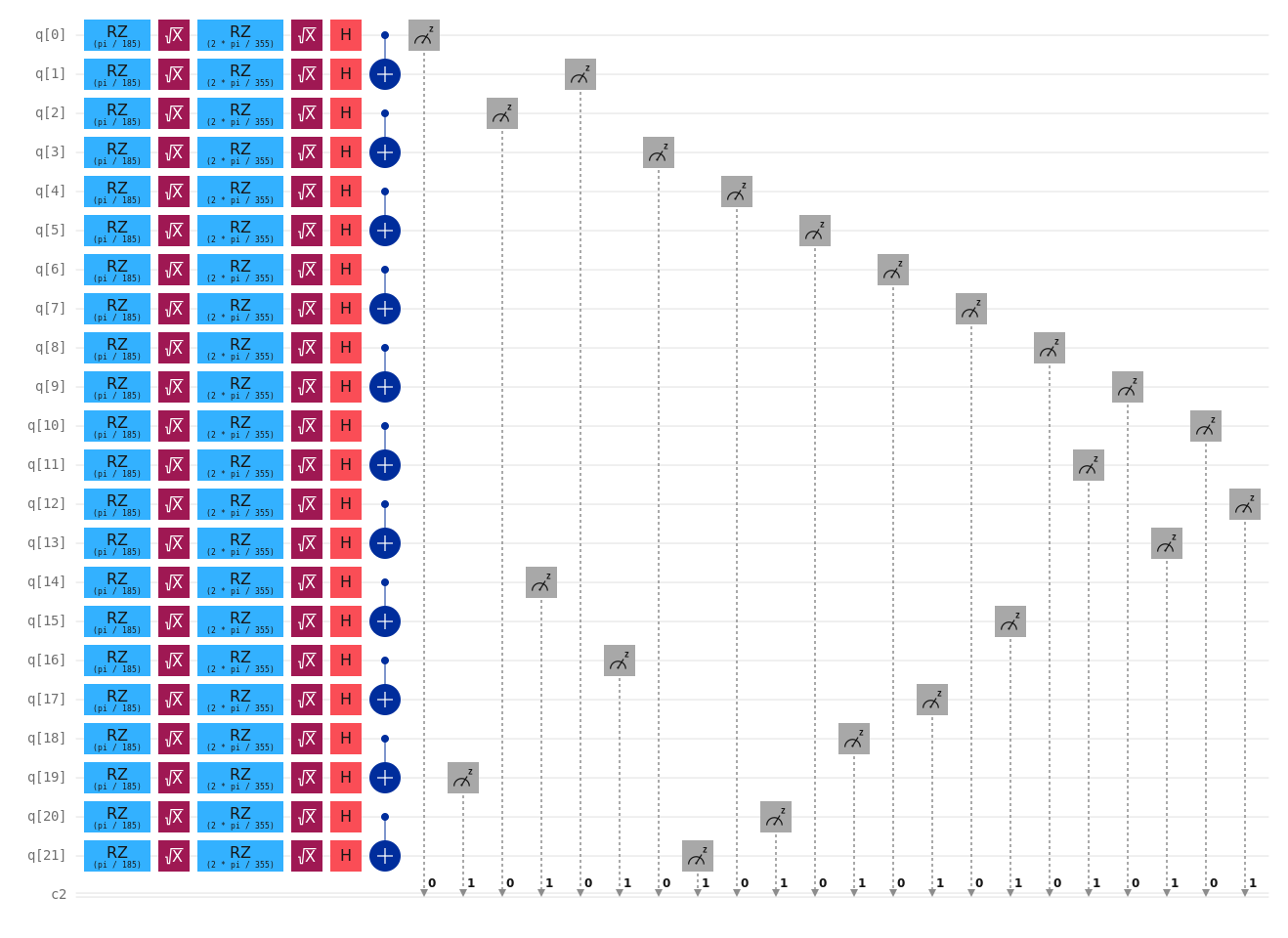}  % Change filename
\caption{Quantum circuits implementing Jaynes–Cummings-inspired gate operations enable scalable modeling of quantum dynamics across multi-qubit systems.}
\end{figure*}

\section{Hamiltonian of Longitudinal Ising Model}
\vspace{2em}
The classical Ising model describes a system of spins that can point either up or down. These spins are arranged in a line (or lattice) and interact with their nearest neighbors. The energy of the system depends on whether neighboring spins are aligned or opposite and whether there's an external magnetic field acting on them.To move from the classical to the quantum Ising model, we make one key change:
Instead of treating spins as just classical up/down values, we treat them as quantum objects—specifically, using quantum spin operators. These quantum spins can exist in superpositions and their behavior follows the rules of quantum mechanics.

The \textbf{Ising model} is a cornerstone of statistical and quantum mechanics, widely used to study spin systems and phase transitions \cite{b37}. In its quantum version, particularly the \textit{longitudinal quantum Ising model}, spins interact with their nearest neighbors along the $z$-axis and may be influenced by an external magnetic field. Unlike the classical version where spins are treated as binary variables, the quantum model describes them using Pauli spin operators. This shift to operator-based formalism allows for a more accurate and powerful representation of quantum phenomena such as superposition and entanglement. As a result, the longitudinal Ising model becomes a useful tool for modeling realistic quantum systems, including arrays of qubits in quantum computers \cite{b38}.

Let us consider a quantum spin chain where each spin interacts only with its nearest neighbor. The energy of interaction between the \(l\) the and \((l+1)\)spins is given by
\begin{equation}
E_{l,l+1} = J_l \sigma_l^z \sigma_{l+1}^z,
\end{equation}
where \( \sigma_l^z \) and \( \sigma_{l+1}^z \) are the Pauli-Z operators corresponding to the spin orientation along the z-axis for the respective spins and \(J_l\) is the coupling constant for that pair. This constant determines the strength and nature of the interaction: a positive \(J_l\) tends to align spins in the same direction, whereas a negative \(J_l\) favors opposite alignments. This type of interaction is a key component in the Ising model, which is fundamental for studying magnetic systems and quantum correlations.

To describe the total interaction energy of a quantum spin chain, where each spin interacts only with its nearest neighbor, we sum over all adjacent spin pairs:
\begin{equation}
H_{\text{int}} = \sum_{l=1}^{N-1} J_l \sigma_l^z \sigma_{l+1}^z,
\end{equation}
where \( J_l \) is the coupling constant between the \( l \)-th and \( (l+1) \)-th spins and \( \sigma_l^z \), \( \sigma_{l+1}^z \) are the Pauli-Z operators acting on those spins.

Each spin is also subjected to a magnetic field along the $z$-axis. The Hamiltonian for this magnetic field interaction is:
\begin{equation}
H_{\text{ext}} = - \sum_{l=1}^{N} h_l \sigma_l^z,
\end{equation}
where \( h_l \) is the magnetic field strength at site \( l \).

Combining both the magnetic field and spin-spin interaction terms, the full Hamiltonian for the longitudinal Ising model is:

\begin{equation}
H_{\text{LIsing}} = H_\text{ext} + H_{\text{int}}
\end{equation}
Substituting Eqns. (12) and (13) in Eqn (14) results,
\begin{equation}
H_{\text{LIsing}} = - \sum_{l=1}^{N} h_l \sigma_l^z + \sum_{l=1}^{N-1} J_l \sigma_l^z \sigma_{l+1}^z.
\end{equation}

This Hamiltonian model is important phenomena in condensed matter, quantum magnetism and quantum computing.
Equation (15) specifies the total Hamiltonian of a one-dimensional quantum spin chain governed by the Ising model without an external magnetic field. It sums over all pairs of neighboring spins, where each pair contributes an energy term of the form \( J_l \sigma_l^z \sigma_{l+1}^z \). Here, \( \sigma_l^z \) and \( \sigma_{l+1}^z \) are Pauli-Z matrices acting on the \(l\)-th and \((l+1)\)-th spins respectively, describing their spin states along the z-axis. The coupling constant \( J_l \) quantifies how strongly each neighboring pair interacts. The sign and magnitude of \( J_l \) determine whether the interaction tends to align the spins in the same direction (ferromagnetic, \( J_l > 0 \)) or opposite directions (antiferromagnetic, \( J_l < 0 \)). By summing over all \( l \), the equation captures the cumulative interaction energy for the entire spin chain, making it a fundamental model for analyzing collective quantum behavior in many-body systems\cite{b39}. This expression represents the full Hamiltonian for a 1D Ising chain without an external magnetic field, capturing the collective behavior of spins through pairwise interactions. 

In our exploration of the Ising model, we introduce a dedicated quantum circuit architecture designed to probe the system’s frequency dynamics, fidelity and inherent error correction capabilities. This endeavor centers on constructing a 22-qubit quantum circuit that faithfully embodies the structure of the Ising model Hamiltonian, as depicted in Fig. 2. The circuit commences with a layer of Hadamard (H) gates, which prepare each qubit in an equal superposition—setting the stage for quantum parallelism. This is followed by an orchestrated application of phase (P) gates and rotation around the z-axis (Rz) gates, each introducing local phase shifts that encode spin interactions specific to the Ising Hamiltonian. Crucially, the backbone of entanglement is constructed through controlled-NOT (CNOT) gates, applied between adjacent qubits to mimic nearest-neighbor couplings and generate the entangled states necessary for capturing collective quantum behavior. The sequence of gates, inspired by the Ising model’s theoretical underpinnings, allows the simulation of complex many-body dynamics on a quantum substrate. Fig. 2 illustrates this evolution, horizontal colored bars denote time-evolving single-qubit operations, while the vertical connectors signify entangling operations that link qubit pairs across the lattice. This layered visual conveys not only the logic of the circuit but also the propagation of quantum correlations essential to simulating spin chain behavior. By structuring the circuit in this manner, we facilitate the investigation of quantum state evolution, identify signatures of quantum phase transitions and assess algorithmic robustness against decoherence and gate imperfections. The implementation also serves as a testbed for refining error mitigation strategies in near-term quantum devices, providing a bridge between theoretical models and their physical realizations.

\begin{figure*}
\centering
\includegraphics[width=1\linewidth]{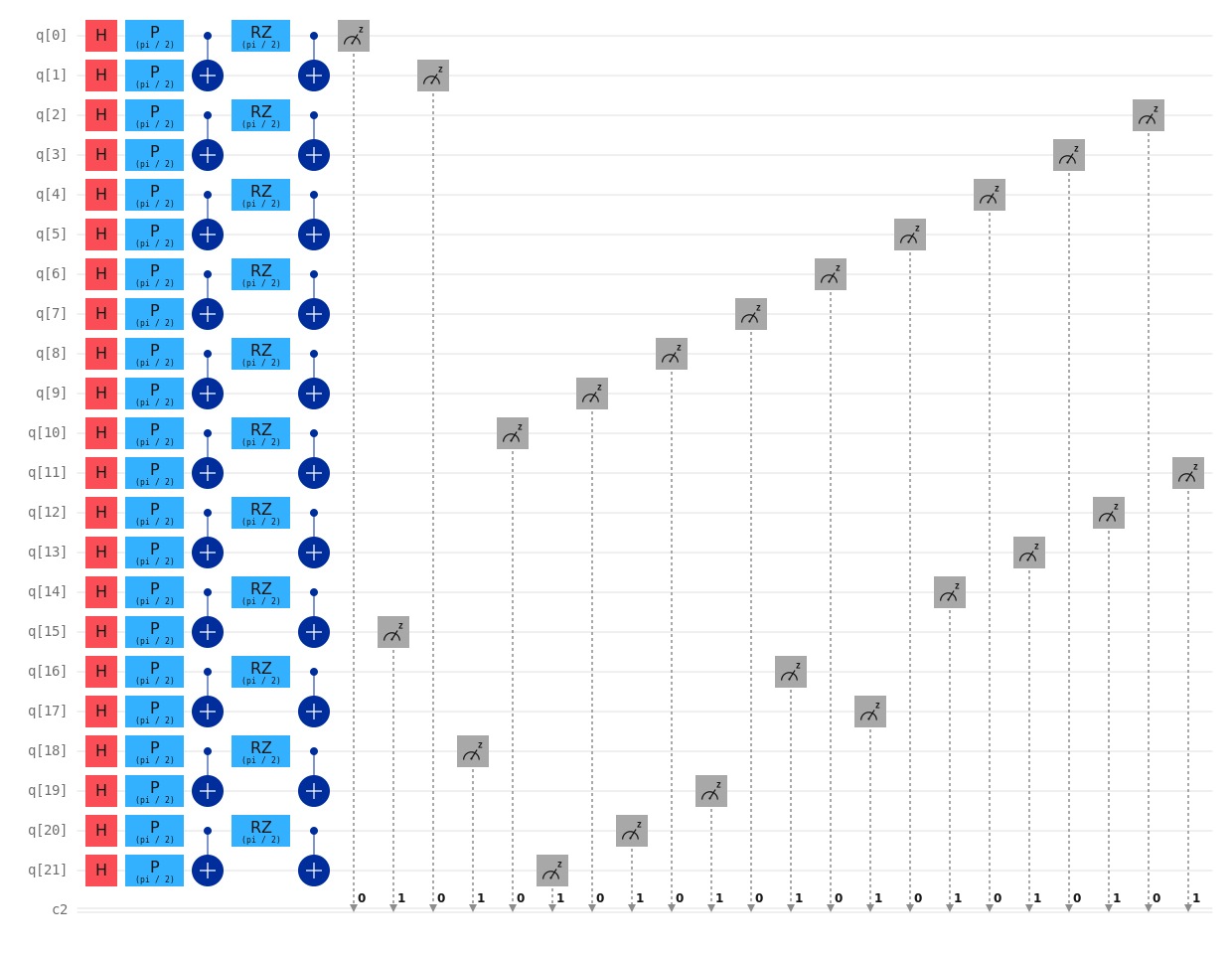}  % Change filename
\caption{Quantum circuits designed with Ising-inspired gate structures enable scalable simulation of spin-chain dynamics, allowing detailed exploration of quantum correlations, many-body interactions, and critical phenomena across multi-qubit systems.} 
\end{figure*}

We employ the \textbf{Brisbane quantum processor} \cite{b40}, recognized as one of the most advanced third-generation quantum chips, owing to its superior speed and qubit manipulation capabilities. Its architecture is particularly optimized for rapid quantum operations, making it an ideal platform for simulating complex quantum systems with enhanced execution efficiency.
In our study, the dynamics of superconducting qubits are explored through precise control and readout techniques, allowing us to probe the evolution of quantum states with minimal latency. Leveraging IBM’s simulation environment, the implementation of our model on the Brisbane processor enables the execution of up to 18,000 quantum operations with minimal reliance on error correction protocols. This capacity not only underscores the processor’s robustness but also illustrates its suitability for running high-depth quantum circuits with remarkable stability and performance.

Figure 3 shows the schematic layout of IBM’s 127-qubit Brisbane processor, where each circle represents a qubit and connecting lines indicate allowed two-qubit interactions which expalains the coupling map, that affects how multi-qubit gates like CNOT are implemented. Such a layout is crucial for optimizing circuit performance on real hardware.

\begin{figure*}[htp]
    \centering
    \includegraphics[width=0.5\linewidth]{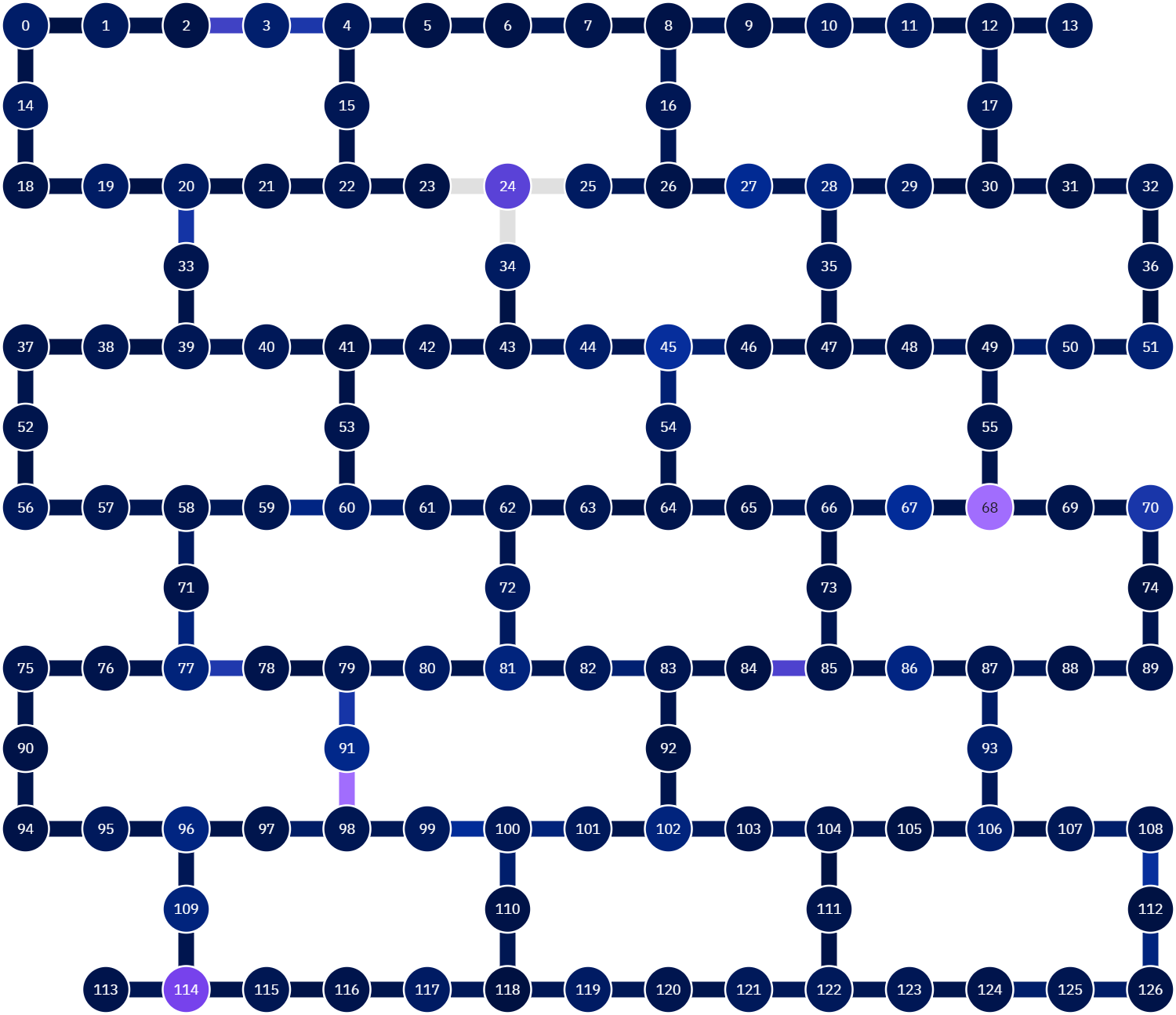}  % Change filename
    \caption{The schematic illustration of IBM's 127-qubit Brisbane processor demonstrates that each circle represents a qubit and the lines connecting them indicate how two qubits interact.
}
\end{figure*}

Superconducting qubits are widely regarded for their precision and stability, making them ideal candidates for scalable quantum computing. To achieve optimal performance and accuracy, these systems typically operate at frequencies between 4 GHz and 6 GHz, a range carefully chosen to balance coherence times with control fidelity. Within this spectrum, microwave signals play a pivotal role—serving as the primary medium through which researchers manipulate and probe quantum states. These signals, often tuned between 5 and 8 GHz, align well with the natural transition frequencies of superconducting qubits, enabling precise state preparation, manipulation and measurement. This frequency window not only ensures minimal interference and signal degradation but also allows for highly accurate and stable characterization of quantum behavior, reinforcing the reliability of superconducting platforms in experimental and applied quantum research. Figure 4 shows two bar graphs that indicate how two alternative qubit interaction models—Jaynes–Cummings (left) and Ising (right) affect the frequency distribution of recorded two-qubit quantum states (00, 01, 10 and 11).  The observed frequencies of downward trend from 00 to 11 in the Jaynes–Cummings model. This behavior stems from the dispersive coupling between qubits and resonators, which causes the energy levels to shift and spread out as more qubits transition into the excited state. As more qubits get into the excited state, the energy levels change, making it less likely that they will be measured in higher-excitation states.  In contrast, the Ising model displays an ascending frequency pattern shows that from 00 to 11. This arises due to  $Z$–$Z$ coupling, which means that the interaction energy increases as more qubits are in the $\ket{1}$ state. By multiplying the number of trials with the potential outcomes of the two quantum states, we compute the resulting frequencies. This process allows us to statistically map the behavior of the quantum system under repeated measurements. The primary objective is to generate a spectrum of frequency values corresponding to various quantum state configurations. By systematically evaluating the resulting frequencies, we aim to assess their consistency and accuracy, thereby verifying the reliability of the quantum circuit in capturing the underlying quantum dynamics.

\begin{figure*}
    \centering
    \includegraphics[width=0.45\linewidth]{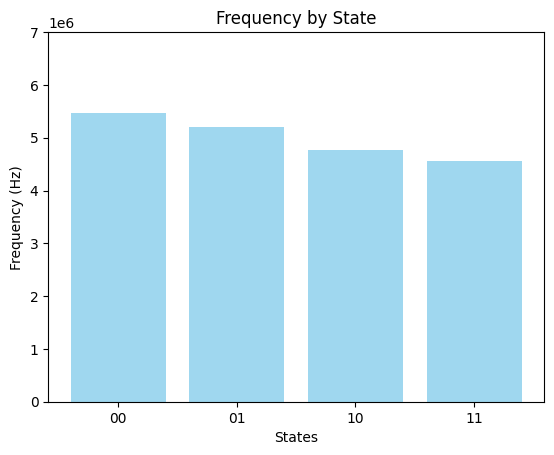} \hfill
    \includegraphics[width=0.45\linewidth]{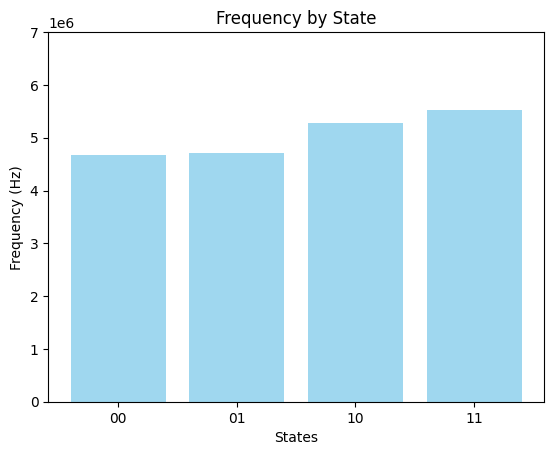}
    \caption{
\textbf{(a)} The bar chart on the left illustrates the frequency distribution of quantum states for a 22-qubit system with  \( 2.98 \times 10^7 \). The y-axis indicates the frequency (in Hz) of each quantum state.
\textbf{(b)} The bar chart on the right displays the frequency distribution of the same quantum states under the \textbf{Ising Model}, also for a 22-qubit system with \( 2.98 \times 10^7 \) shots. This comparison highlights how different quantum interaction models influence the observed distribution of states under the same simulation conditions.
    }
    \label{fig:frequency_comparison}
\end{figure*}
In general, adding more quantum gates while keeping the number of qubits low but enough tends to improve fidelity [41], as long as the superconducting qubits stay in their best operational states.  Figure 5 shows that if the number of gates goes up with the number of qubits, the fidelity value may go down, which means that the error rate goes up.

\begin{figure}[htp]
    \centering
    \includegraphics[width=0.95\linewidth]{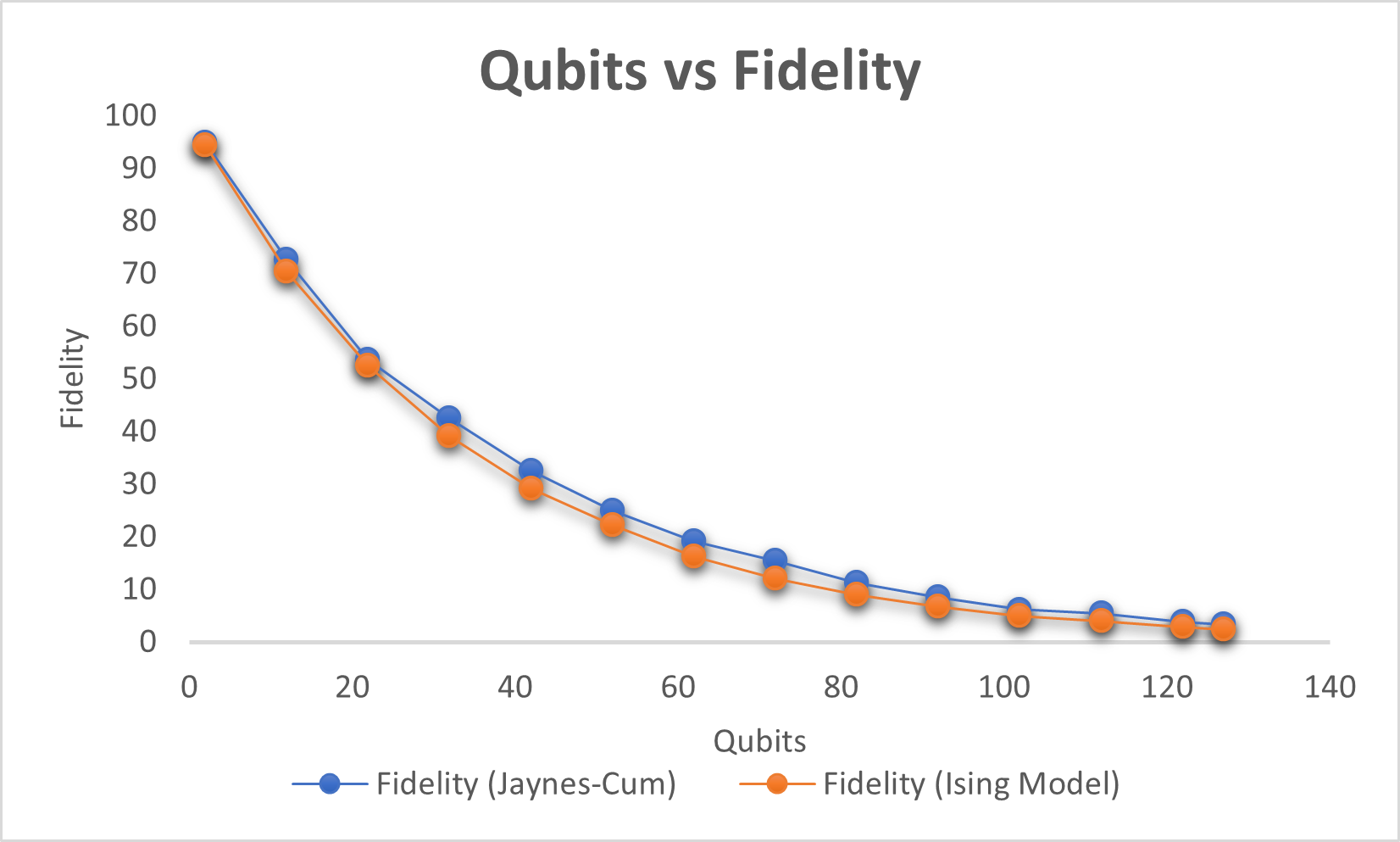}  % Change filename
    \caption{Fidelity comparison between the Jaynes–Cummings and Longitudinal Ising models across varying qubit counts. The plot illustrates the fidelity degradation from 2 to 127 qubits, highlighting the performance scalability of each model. The Jaynes–Cummings model demonstrates superior resilience to noise and decoherence, while the Ising model reveals limitations under large-scale quantum simulation.}
\end{figure}

This divergence between the two models reflects fundamental differences in how qubit interactions shape quantum state probabilities and energy landscapes. Understanding the behavior of superconducting qubits under various computing loads is made easier by these simulations. In high-speed quantum computations, fidelity is a crucial criterion for assessing the stability and quality of superconducting qubits.

Fig. 6 illustrates how increase in bit errors within the quantum circuit with  diminishes the output of the qubits.  This behavior can be efficiently detected and evaluated in IBM's quantum technologies. Classical bits are frequently utilized when implementing frequencies in a limited number of classical systems.  By establishing the values of the classical bits, the whole frequency may be delineated to enhance comprehension of the behavior of these bits. This technique enables enhanced precision in quantifying key quantum properties. It shows how closely the desired state and the actual state of a qubit match and it is essential for assessing how well quantum error correction works, especially in
connection with the quantity of qubits used. Quantum operators are implemented within circuits to analyze and demonstrate these phenomena. 
Error correction demands grow with increasing qubit count in both interaction models, underscoring the inherent fragility of larger quantum systems. Notably, the Ising model exhibits a slightly higher requirement for error correction compared to the Jaynes–Cummings model, reflecting the stronger inter-qubit correlations and more complex entanglement structures involved. This trend highlights the critical importance of robust error mitigation strategies as system size scales, ensuring the fidelity and stability of quantum computations in both theoretical simulations and real hardware implementations.
\begin{figure}[htp]
    \centering
    \includegraphics[width=0.95\linewidth]{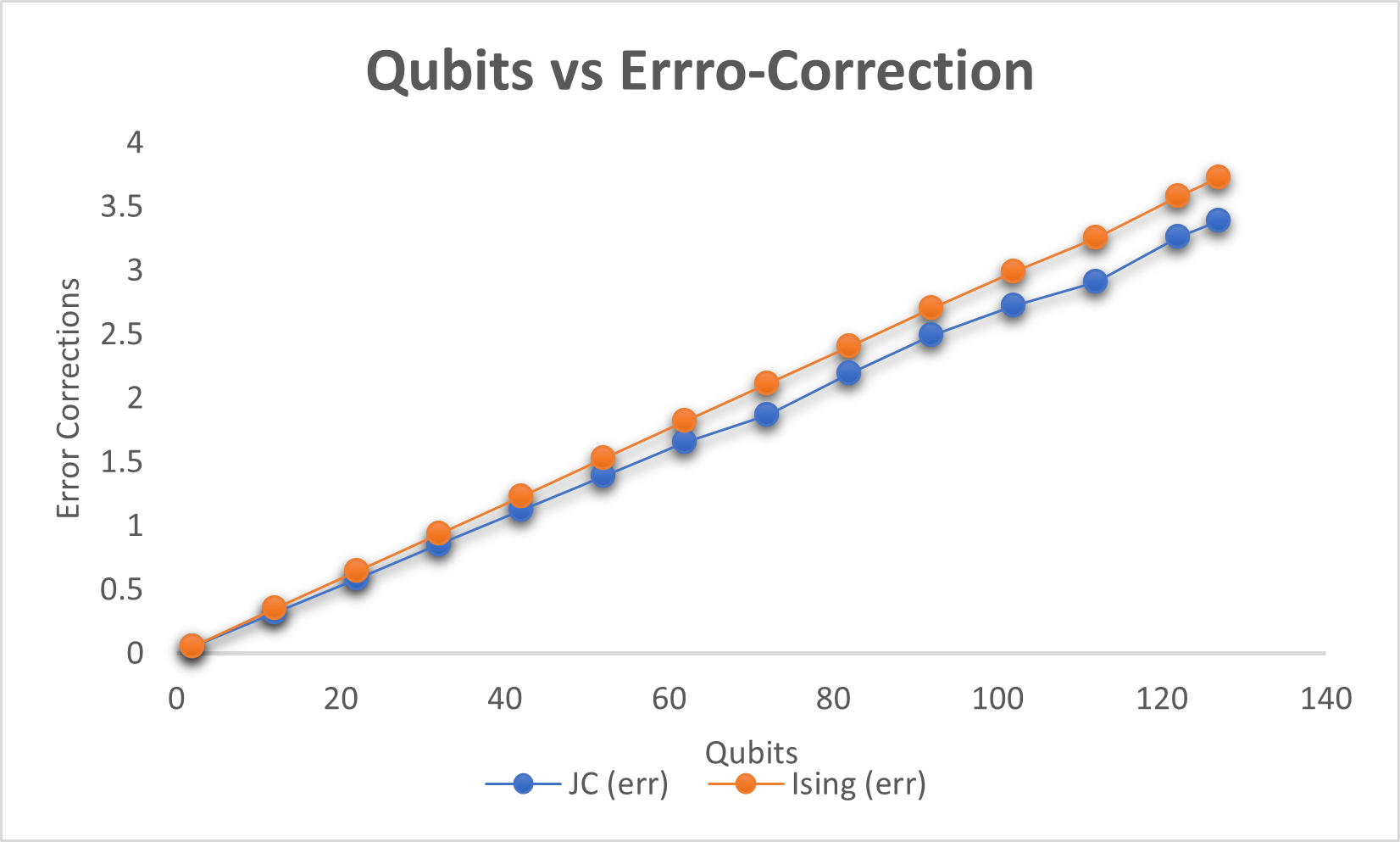}  % Change filename
    \caption{Error correction performance across quantum systems of increasing size, from 2 to 127 qubits. The graph reveals a rising trend in cumulative errors with larger qubit counts, emphasizing the critical need for advanced error mitigation and fault-tolerant strategies in scalable quantum computing architectures.}
\end{figure}

\begin{figure}[htp]
    \centering
    \includegraphics[width=0.95\linewidth]{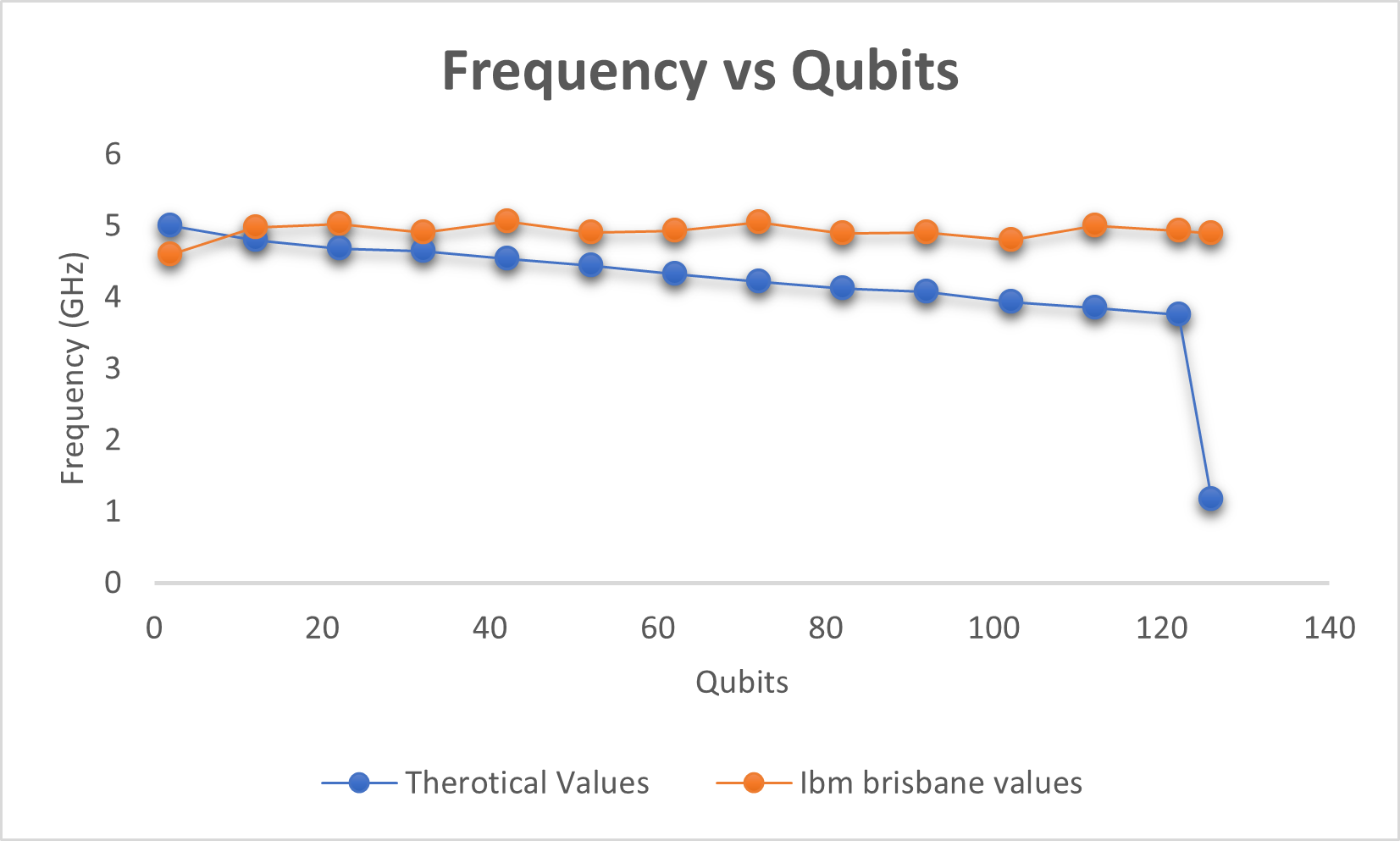}  % Change filename
    \caption{Comparison of qubit frequencies across systems ranging from 2 to 127 qubits, based on theoretical predictions and IBM Brisbane hardware data. While theoretical models indicate a gradual frequency decline followed by a sharp drop near 126 qubits, the IBM Brisbane hardware maintains stable qubit frequencies between 4.9 and 5.0 GHz, demonstrating superior frequency control and calibration stability in large-scale quantum systems.}
\end{figure}

\begin{table*}[ht]
\centering
\begin{tabular}{|p{1cm}|p{2cm}|p{1.5cm}|p{1.5cm}|p{1.5cm}|p{1.5cm}|p{1.5cm}|}
\hline
\multirow{2}{*}{\textbf{Qubits}} & \multicolumn{2}{c|}{\textbf{Frequency (GHz)}} &
\multicolumn{2}{c|}{\textbf{Fidelity}} &
\multicolumn{2}{c|}{\textbf{Error Correction}} \\
\cline{2-7}
& \textbf{Theoretical} & \textbf{IBM Brisbane} & \textbf{Jaynes-Cumming Model} & \textbf{Ising Model} & \textbf{Jaynes-Cumming Model} & \textbf{Ising Model} \\
\hline
2   & 5.00 & 4.60 & 94.79 & 94.29 & 0.0534 & 0.0587 \\
12  & 4.80 & 4.98 & 72.55 & 70.29 & 0.3208 & 0.3525 \\
22  & 4.68 & 5.03 & 53.53 & 52.40 & 0.5882 & 0.6462 \\
32  & 4.65 & 4.91 & 42.50 & 39.06 & 0.8556 & 0.9400 \\
42  & 4.54 & 5.06 & 32.53 & 29.11 & 1.1229 & 1.2337 \\
52  & 4.44 & 4.91 & 24.89 & 22.14 & 1.3903 & 1.5275 \\
62  & 4.33 & 4.93 & 19.05 & 16.18 & 1.6575 & 1.8212 \\
72  & 4.22 & 5.05 & 15.41 & 12.06 & 1.8696 & 2.1153 \\
82  & 4.12 & 4.89 & 11.16 & 8.99  & 2.1924 & 2.4087 \\
92  & 4.03 & 4.91 & 8.54  & 6.70  & 2.4958 & 2.7025 \\
102 & 3.94 & 4.80 & 6.53  & 4.99  & 2.7272 & 2.9962 \\
112 & 3.85 & 5.00 & 5.45  & 3.91  & 2.9082 & 3.2395 \\
122 & 3.76 & 4.93 & 3.83  & 2.77  & 3.2619 & 3.5837 \\
126 & 1.17 & 4.90 & 3.36  & 2.39  & 3.3923 & 3.7305 \\
\hline
\end{tabular}
\caption{System performance trends as a function of qubit number, ranging from 2 to 127. The graph captures the interplay between fidelity, frequency, and error rates, revealing that as qubit count increases, fidelity and frequency generally decline, while error accumulation intensifies. The comparison integrates theoretical predictions with empirical data from IBM’s Brisbane quantum processor, highlighting the hardware’s relative stability and scalability challenges.}
\end{table*}
To overcome hardware-imposed shot limitations, we employ a strategy of aggregating the outcomes of multiple low-shot executions on IBM Quantum systems, as illustrated in Figure 7. This approach enables the collection of highly accurate measurement data without exceeding the permissible number of shots per run, making it particularly valuable for probing larger quantum systems.
Figure 7 reveals a striking trend: as the number of qubits increases, the system's operating frequency undergoes a significant decline. While smaller circuits—comprising fewer qubits—maintain stability within the 4–5 GHz range, the frequency plummets to nearly 1 GHz when scaling up to 127 qubits. This sharp drop indicates the growing challenge of preserving coherent control and synchronization as system complexity escalates. The comparison presented in the figure also underscores how different underlying Hamiltonians influence quantum state evolution and readout probabilities. These observed dynamics are crucial for the development of high-fidelity quantum gates, the design of scalable quantum algorithms and the calibration of realistic circuit models. 
Figure 7 illustrates a resource-efficient technique employed in IBM Quantum systems, where multiple operations are executed with fewer shots and the outcomes are subsequently aggregated. This approach enables the extraction of highly accurate measurement data while remaining within the shot limitations imposed by the hardware. It proves especially effective when scaling quantum circuits for large qubit systems. A clear trend emerges inFigure 7 shows that the system's operating frequency goes up as the number of qubits goes up .While smaller systems exhibit stability within the 4–5 GHz range, the frequency sharply drops to approximately 1 GHz when the qubit count reaches 127. This decline highlights the increasing difficulty in maintaining precise control and synchronization in large-scale quantum architectures. Moreover, the comparison in the figure emphasizes how the system’s Hamiltonian significantly shapes both the quantum state evolution and the readout probabilities. Understanding these frequency dynamics is essential for the development of high-fidelity quantum gates, the design of scalable quantum algorithms,the accurate modeling, simulation and calibration of complex quantum circuits. We take into consideration three main causes of error when assessing the overall fidelity of a quantum circuit running on IBM's 127-qubit Brisbane processor: Measurement error, two-qubit gate errors, and single-qubit gate defects Each of these defect types contributes uniquely to the degradation of circuit performance. The total error probability is calculated as follows
\newline
\newline
Single-qubit Error :

Assuming 12 single-qubit gates are applied and each gate has an average error rate of \( 2.596 \times 10^{-4} \),
\[
\text{Error}_{\text{single}} = 12 \times (2.596 \times 10^{-4}) = 0.002755.
\]
Two-qubit Error :

Assuming 2 CNOT gates with an average error rate of \( 6.560 \times 10^{-3} \),
\[
\text{Error}_{\text{two}} = 2 \times (6.560 \times 10^{-3}) = 0.01389.
\]
Measurement Error :

The average measurement error is given as:
\[
\text{Error}_{\text{Measurement}} = 0.0534.
\]
Total Error and Fidelity:

Let \( E_S, E_T, E_M \) represent the single-qubit, two-qubit and measurement errors respectively. The overall fidelity \( F' \) is computed as,

\[F' = e^{-(E_S + E_T + E_M)}\]

Substituting the values,

\[F' = e^{-(0.002569 + 0.06560 + 0.01587)} = e^{-0.0534}\]
\[F' \approx 0.9479 \approx {94.79\%}\]
This signifies that the total circuit fidelity on the IBM Brisbane processor for the specified configuration is roughly 94.79\%, which is regarded as rather high in near-term quantum devices (NISQ era) \cite{b42}.  In superconducting qubits, an augmentation in the quantity of qubits is associated with an escalation in error correction demands.   As a result, the fidelity value decreases.   This suggests that as the quantity of qubits increases, the overall fidelity declines.  In a two-qubit system, fidelity is improved with little error correction.   However, when the quantity of qubits escalates, error correction requires supplementary gates and repeated cycles, leading to the proliferation of errors inside quantum states.
\section{Conclusion }
\vspace{1em}
In conclusion, this study provides a comprehensive investigation into the behavior and performance of superconducting quantum circuits utilizing transmon qubits on IBM’s Brisbane quantum processor. By integrating the Jaynes–Cummings interaction and the Longitudinal Ising model into our simulations, we unveiled distinct strengths and limitations associated with each framework—highlighting the Jaynes–Cummings model’s superiority in preserving fidelity and frequency stability across larger qubit architectures. Notably, the sharp decline in performance beyond a few qubits underscores the persistent challenges posed by gate infidelity, decoherence and measurement noise in current quantum hardware. Through rigorous modeling, we demonstrated how fidelity deteriorates with increasing system size, thereby drawing attention to the delicate trade-off between scalability and quantum coherence. The Longitudinal Ising model, though less robust in large-scale implementations, proved insightful in exploring complex many-body dynamics and condensed matter analogs. These findings collectively stress the critical need for innovations in superconducting circuit design, improved qubit connectivity, and the integration of more resilient error-correction protocols. As the field advances, addressing these bottlenecks will be vital for achieving practical, large-scale quantum computing. Superconducting platforms, with their solid-state reliability and strong industry support, remain at the forefront of this pursuit. Continued collaboration between theory, experiment, and engineering will be pivotal in transcending current limitations and realizing the full potential of quantum technologies.

\section*{Acknowledgments}
\vspace{1em}
S. Saravana Veni acknowledges Amrita Vishwa Vidyapeetham, Coimbatore, for the funding provided under the Amrita Seed Grant, Israel (File Number: ASG2022141).

\end{document}